# MgB$_2$ single crystals: high pressure growth and anisotropic properties

*J.Karpinski[1], M.Angst[1], J.Jun[1], S.M.Kazakov[1], R.Puzniak[2], A.Wisniewski[2], J.Roos[3], H.Keller[3], A. Perucchi[1], L. Degiorgi[1], M.Eskildsen[4], P.Bordet[5], L.Vinnikov[6], A.Mironov[7], [1]Solid State Physics Laboratory ETH-Zürich, [2]Institute of Physics, Polish Academy of Sciences, Warsaw, [3]Physik-Institut, Universität Zürich, [4]DPMC, University of Geneva, [5]Laboratoire de Cristallographie C.N.R.S. Grenoble, [6]Ames Laboratory, US Department of Energy, Ames, Iowa, [7]Chemical Department, Moscow State University

**Abstract**

Single crystals of MgB$_2$ with a size up to 1.5x0.9x0.2 mm$^3$ have been grown with a high pressure cubic anvil technique. The crystal growth process is very peculiar and involves an intermediate nitride, namely MgNB$_9$. Single crystals of BN and MgB$_2$ grow simultaneously by idth of ~0.5 K. The high quality of the crystals allowed the accurate determination of magnetic, transport (electric and heat) and optical properties as well as scanning tunnelling spectroscopy (STS) and decoration studies. Investigations of crystals with torque magnetometry show that $H_{c2}^{//c}$ for high quality crystals is very low (24 kOe at 15 K) and saturates with decreasing temperature, while $H_{c2}^{//ab}$ increases up to 140 kOe at 15 K. The upper critical field anisotropy $\gamma = H_{c2}^{//ab}/H_{c2}^{//c}$ was found to be temperature dependent (decreasing from $\gamma \cong 6$ at 15 K to 2.8 at 35 K). The effective anisotropy $\gamma_{\text{eff}}$, as calculated from reversible torque data near $T_c$, is field dea peritectic decomposition of MgNB$_9$. Magnetic measurements with SQUID magnetometry in fields of 1-5 Oe show sharp transitions to the superconducting state at 37-38.6 K with wpendent (increasing roughly linearly from $\gamma_{\text{eff}} \cong 2$ in zero field to 3.7 in 10 kOe). The temperature and field dependence of the anisotropy can be related to the double gap structure of MgB$_2$ with a large two-dimensional gap and small three-dimensional gap, the latter of which is rapidly suppressed in a magnetic field. Torque magnetometry investigations also show a pronounced peak effect, which indicates an order-disorder phase transition of vortex matter. Decoration experiments and STS visualise a hexagonal vortex lattice. STS spectra in zero field evidence two gaps 3 meV and 6 meV with a weight depending on the tunnelling direction. Magneto-optic investigations in the far infrared region with $H//c$ show a clear signature of the smaller of the two superconducting gaps, completely disappearing only in fields higher than $H_{c2}^{//c}$.

*Corresponding author: Fax:+4116331072, e-mail: karpinski@solid.phys.ethz.ch

## 1. Introduction

The recent discovery of superconductivity at 39K in MgB$_2$ by Akimitsu *et al.* [1] has stimulated world wide excitement. The transition temperature is above or on the limit suggested theoretically for BCS phonon mediated superconductivity. One of the questions with regards to potential applications is whether this new superconductor resembles a high $T_c$ cuprate superconductor or low $T_c$ metallic superconductor in terms of its current carrying characteristics in applied magnetic fields.

At the beginning of MgB$_2$ studies, the majority of investigations were performed on polycrystalline samples. In order to understand the intrinsic properties of this structurally highly anisotropic compound, investigations of anisotropic parameters should be performed on single crystals. Some intrinsic properties are now better understood and to our knowledge well established [2-4]:
1. There are two bands, 3D and 2D which take part in the superconductivity.
2. There are two superconducting gaps, smaller connected with 3D $\pi$ band and larger connected with 2D $\sigma$ band.
3. As one can see in tunneling experiment in the *c*-direction, smaller $\pi$ gap is suppressed at relatively low magnetic field, second, larger $\sigma$ gap survives at much higher fields.

A torque study [5] reviewed in this paper found the upper critical field anisotropy $\gamma=H_{c2}^{//ab}/H_{c2}^{//c}$ to be temperature dependent, increasing from about 2 at temperatures close to $T_c$ to about 6 at low temperatures. Other measurements on crystals grown as described below [6,7] and on crystals obtained with other methods [8,9] confirm this temperature dependence, although exact numbers vary between different reports [5-9].

In our torque studies on single crystals, we show that the anisotropic Ginzburg-Landau theory with a temperature and the field independent effective mass anisotropy does not work for $MgB_2$. Based on the results of the magnetic torque measurements we propose a $H,T$ phase diagram for $MgB_2$ compound.

We will also discuss additional measurements on the crystals we have grown, observing directly energy gaps (STS, magneto-optics) and visualizing the vortex lattice (STS, decoration) in $MgB_2$.

## 2. Crystal growth

Since $MgB_2$ melts non-congruently, it is not possible to grow crystals from a stoichiometric melt. Therefore, we decided to apply a high temperature solution growth method. The solubility of $MgB_2$ in Mg is extremely low at temperatures below the boiling temperature of Mg (1107$^o$C) at normal pressure (Fig. 1a)), therefore crystals have to be grown at much higher temperature or by using another solvent. We have tried both ways. At higher temperatures at normal pressure, Mg does not exist as a condensed phase, therefore in order to grow crystals from solution in Mg, pressure has to be increased. Above 1600°C the solubility is high enough, however, the vapor pressure of molten Mg is rather high, above 50 bar. According to the calculated *P-T* phase diagram [10] (shown in Fig.1b)) crystal growth from solution has to be performed in conditions above the boiling line of Mg, which is the border between Liquid+$MgB_2$ and Gas+$MgB_2$.

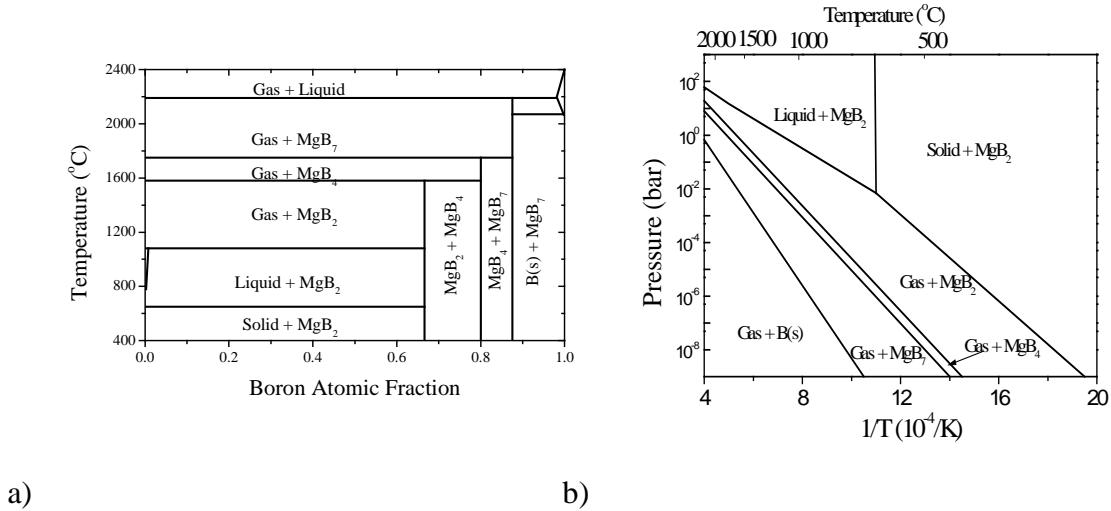

Fig.1. a) Temperature-composition phase diagrams of the Mg-B system under the pressure of 1 bar. b) Pressure-temperature phase diagram for the Mg:B atomic ratio $x_{Mg}/x_B \geq 1/2$. The region of Liquid+$MgB_2$ represents the thermodynamic stability window for the growth of $MgB_2$ from solution in Mg. All data taken from Ref. [10].

We have investigated several methods of high-temperature solution crystal growth:

a) Crystal growth of $MgB_2$ from a solution in Al, Mg, Cu, and mixtures of these metals, at high temperatures up to 1700°C at Ar pressure 1bar<$P_{Ar}$<14kbar.
b) Crystal growth of $MgB_2$ from a solution in Mg at high temperatures up 2000°C, using a cubic anvil technique with a solid pressure medium (10<P<35 kbar).

There are several major problems to be overcome:

(I)  The reactivity of the crucible material. At high temperatures above 1000°C molten Mg is very aggressive towards all materials and destroys the crucible after several hours.
(II) All metals used as a solvent form mixed compounds with Mg or $MgB_2$, which makes crystal growth of pure $MgB_2$ impossible from any solvent other than Mg. Using Al-Mg flux, we can grow mixed $Mg_xAl_{1-x}B_2$ crystals.
(III) The solubility of $MgB_2$ in Mg is very small at temperatures below 1600°C. At 1600°C the partial pressure of Mg vapors above molten Mg is of the order of 50 bar.
(IV) $MgB_2$ decomposes at high temperatures, above 1000°C, at normal pressure.

## 3. Crystal growth experiments

(a) In order to grow crystals in an argon atmosphere, we have employed our gas pressure apparatus, used previously for high pressure growth of cuprates [11]. In a molybdenium crucible, we placed a BN crucible, containing Al, Mg, Cu, or a mixture of these metals with B. Experiments have been performed at temperatures from 1000 up to 1700°C at Ar pressure up to 1000 bar. Cu and Al, used as a solvent, form mixed compounds or solid solutions with Mg or $MgB_2$. This makes the growth of pure $MgB_2$ crystals impossible from any solvent other than Mg.

(b) We have also performed crystal growth experiments at an argon pressure of 14 kbar. The BN crucible containing Mg and B was heated up to 1250°C with a temperature gradient of 50°C between colder and warmer part, and kept at the maximum temperature for 10 hours. After this time, no crystals of $MgB_2$ have been found. Growth of pure $MgB_2$ crystals from a Mg solution was not possible in Ar pressure.

(c) In order to grow pure $MgB_2$ crystals, we have applied a cubic anvil technique. The cubic anvil device used in our work is shown in Fig. 2. The pressure transmitting medium is a pyrophylite cube of 22 mm edge size.

Typically, the crystal growth experiments have been performed at a pressure of 30-35 kbar. A mixture of Mg and B was put into a BN container of 6 mm internal diameter and 7 mm length. The temperature was increased during one hour up to the maximum of 1700-1800°C, kept for 1-3 hours and decreased during 1-2 hours. As a result, we obtained a collection of $MgB_2$ and BN crystals sticking together. $MgB_2$ plate like crystals were up to 1.5x0.9x0.2mm$^3$ in size and up to 230 µg in weight. Examples are shown in Fig. 3. The crystals typically have transition temperatures between 37 and 38.6K with a width of 0.4-0.6K. Figure 4 shows a typical dependence of magnetic moment *m* vs temperature *T* of one of the single crystals grown, measured in a field of 2 Oe by SQUID magnetometry. A sharp transition to the superconducting state is seen, with an effective $T_c$ = 38.4 K (onset: 38.6 K) and a transition width (10%-90% criterion) of 0.5 K, indicative of a high quality of the crystal. The relatively large size and good quality of the single crystals allowed for detailed investigations of the intrinsic properties of $MgB_2$ with various methods.

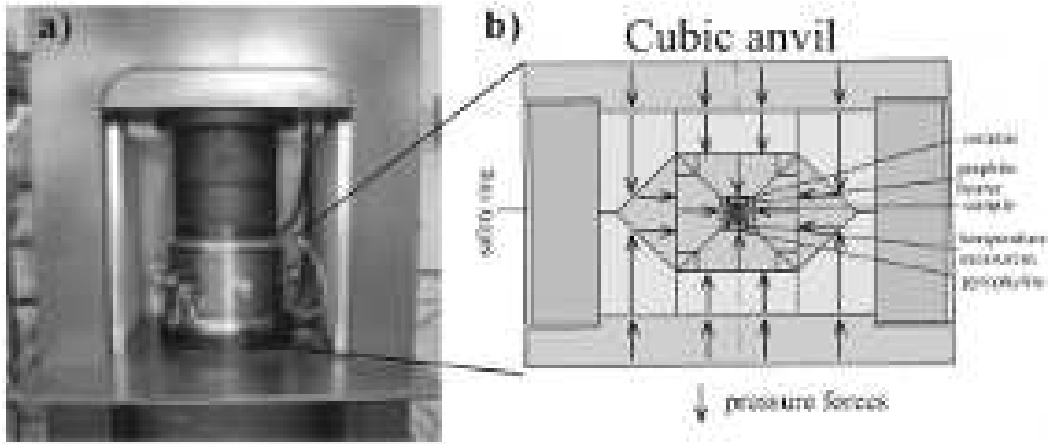

Fig.2. a) Photograph of the cubic anvil device. The force is supplied by the hydraulic press. b) Scheme of the cubic anvil cell. Moveable steel pieces are arranged in such a way as to provide forces pressing from all sides onto the sample in the middle. Heating is provided by passing about 400A through a graphite tube inside the pyrophylite cube. The sample is placed in a BN container inside the graphite heater.

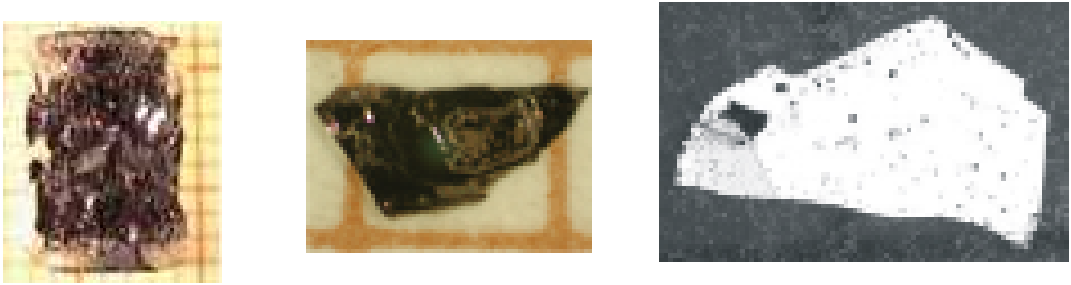

a)                              b)

Fig.3. a) Sample containing $MgB_2$ and BN crystals after synthesis. b) Single crystals of $MgB_2$.

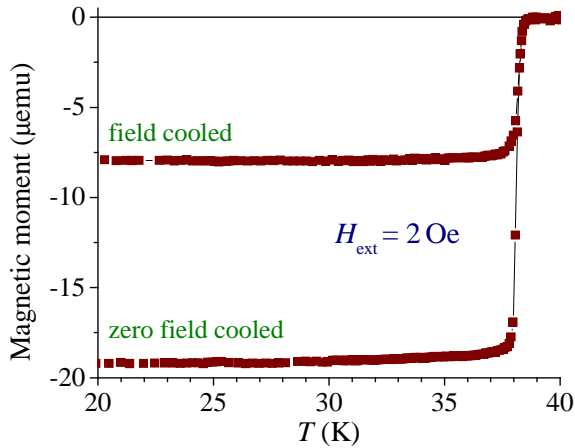

Fig. 4: Magnetic moment vs temperature of a $MgB_2$ single crystal in 2 Oe.

Sometimes, we observed additional hexagonal black crystals, which were involved in the crystal growth process. In some cases, we found $MgB_2$ crystals grown inside of these black crystals (Fig.5a)). In the experiments performed at argon pressure these black crystals were usually found in the crucible. Structural x-ray studies showed that the black hexagonal crystals are a new phase, namely $MgNB_9$. The crystal growth process of $MgB_2$ is not a simple growth from a solution in molten Mg. Rather, crystals grow by a peritectic decomposition of an intermediate nitride phase. $MgB_2$ crystals are the product of a reaction in the ternary Mg-B-N system shown in Fig.5b). The source of nitrogen is the BN crucible, which reacts with molten Mg and B, forming the $MgNB_9$ compound. This compound decomposes at high temperature to $MgB_2$, according to the reaction:

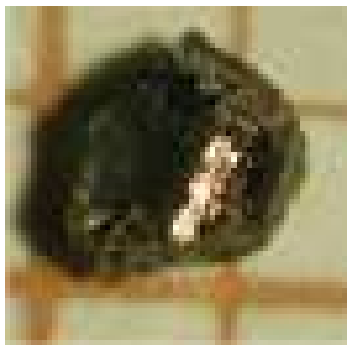
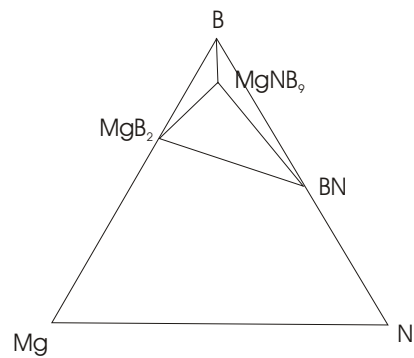

a)  b)
Fig.5. a) Crystal of the new phase $MgNB_9$ with golden $MgB_2$ crystals grown inside.
b) Ternary Mg-B-N system with compounds involved in the crystal growth of $MgB_2$.

$Mg + 8B + BN \rightarrow MgNB_9 \rightarrow MgB_2 + BN + 6B$

The second reaction takes places only at high pressure above 20 kbar. At lower pressure, $MgNB_9$ remained as a final product. Therefore, we have here an example of a rare situation: a pressure induced decomposition of a solid phase. The volume of the products of the decomposition reaction is 1% lower than the volume of the $MgNB_9$ compound, what can induce the decomposition at high pressure conditions. The structure of $MgNB_9$ is shown in Fig.6. It contains two kinds of boron polyhedra separated by Mg and N atoms. The detailed description of the MgNB$_9$ structure is published in Ref.12.

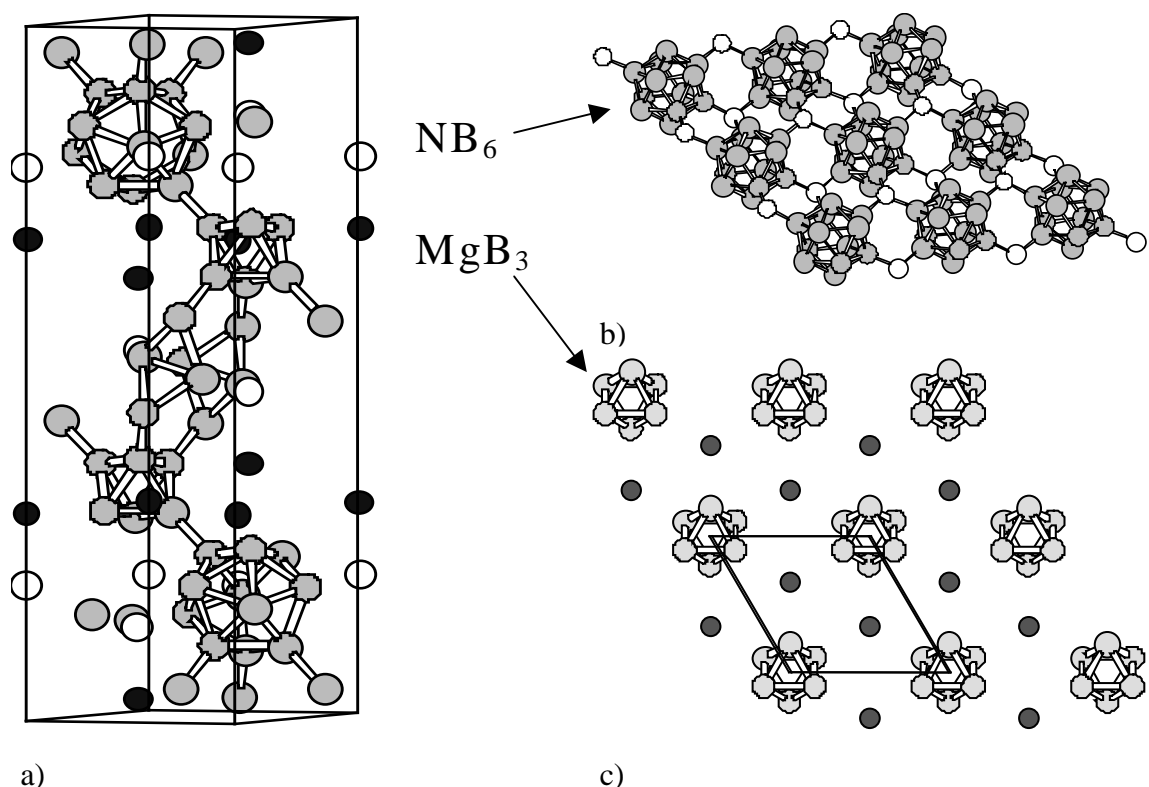

a)                                                       c)

Fig.6. a) Structure of the new compound MgNB$_9$. White circles indicate N, grey B and black Mg atoms. b) The NB$_6$ layer in the MgNB$_9$ structure. c) The MgB$_3$ layer.

## 4. Structural investigations

In order to check the crystal structure of our samples, we carried out x-ray diffraction measurements on single crystals from four different batches for which $T_c$ varied from 34.4K to 38.8K. The crystals were in the form of hexagonal platelets with dimensions between 0.12x0.03 mm$^2$ and 0.4x0.06 mm$^2$. Similar x-ray measurements were applied for all samples. The data collection were carried out on a Nonius Kappa CCD diffractometer, using graphite monochromated AgK$\alpha$ radiation, up to a resolution of $\sin\theta/\lambda \approx 1$, by measuring 2 deg oscillation frames with a sample-to-detector distance of 26 mm. This way, about 3400 reflexions were measured, corresponding to about 99% completeness and a redundancy of over 30. Integrated intensities were extracted with the Eval CCD software [13], a gaussian absorption correction using the crystal dimensions was applied using MaXus [14], and the reflexion merging and structure refinements were carried out with Jana2000 [15].

In all cases, roughly 65 unique reflexions with $\sin\theta / \lambda > 0.2$ were used for structure refinements, and the obtained $R$ and $Rw$ agreement factors were between 1.5 and 2%. The atoms were placed in their standard positions for the AlB$_2$-type structure (Mg (0 0 0), B(2/3 1/3 ½)), with anisotropic displacement parameters (a.d.p.). An isotropic extinction correction (type 1, Lorentzian distribution) was applied [16]. The structural parameters obtained for the four samples (cell parameters and a.d.p.'s) were all equal within two estimated standard deviations (e.s.d.'s). Their values are:

a = 3.085(1) Å, c = 3.518(2)Å, $U_{11}$(Mg) = 0.0045(1)Å$^2$, $U_{33}$(Mg) = 0.0048(2) Å$^2$, $U_{11}$(B) = 0.0034(2) Å$^2$ and $U_{33}$(B) = 0.0044(2) Å$^2$. This agreement between structural data of the crystals from different batches indicates that the changes of $T_c$ observed between crystals from different batches cannot be attributed to detectable structural modifications. Attempting to refine the B atom occupancy lead to values equal to 1 to better than 1% e.s.d. in all cases. However, we should mention that a minor substitution of boron by carbon or nitrogen from the heater or crucible would be almost impossible to detect with x-ray diffraction, because of the small contrast between these elements, and therefore cannot be excluded here.

As also observed by Nishibori et al. [17] and Lee et al. [18], difference fourier maps reveals a remaining electron density connecting the boron atoms, as expected for an sp$^2$ orbital arrangement. A temperature-dependent electron density to investigate fine structural modifications across the superconducting phase transition study is in progress.

### 4. Torque magnetometry investigations

A method particularly well suited to investigate the superconducting state of (anisotropic) single crystals is torque magnetometry. Here, the mechanical torque $\tau = m \times B \approx m \times H$, where $m$ is the sample magnetic moment, $B$ is the induction, and $H$ is the applied magnetic field, is measured, in our case with a miniaturized piezo-resistive technique (the experimental arrangement used is described in Ref. [5]).

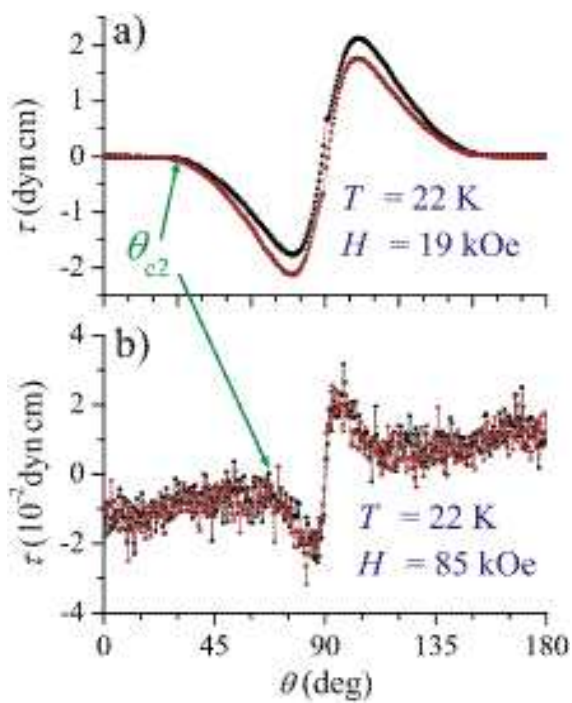

Fig. 7: Torque $\tau$ vs angle $\theta$ between the applied field and the $c$-axis of a MgB$_2$ single crystal, at 22 K in a field of 19 kOe (a) and in a field of 85 kOe (b). $\theta_{c2}$ marks the crossover angle between normal and superconducting states. The data have been antisymmetrized around 90 deg to remove the symmetric part of the background.

In measurements of torque as a function of angle (Fig. 7), a flat part can be seen, where the torque is essentially zero (apart from a small background contribution), indicating that the crystal is in the normal state. Only for field directions close to the direction parallel to the *ab* plane there is a superconducting torque signal. The crossover angle $\theta_{c2}$ between these regions is therefore the angle for which the applied field is the upper critical field. Without any further analysis, it follows immediately from Fig. 7 a) and b) that both 19 kOe and 85 kOe are higher than $H_{c2}^{//c}$, but lower than $H_{c2}^{//ab}$. Therefore, we see that at 22 K, the upper critical field anisotropy $\gamma = H_{c2}^{//ab}/H_{c2}^{//c}$ must be higher than $85/19 \approx 4.47$. Note that the establishment of this lower bound involves neither any upper critical field criterion nor any fitting procedure, but follows straightforwardly from the existence of both flat and non-flat torque angular regions under these conditions. From similar lower and upper bounds at various temperatures it became evident that the $H_{c2}$ anisotropy cannot be independent of temperature.

For a more exact analysis [5], we used a scaling procedure based on the theory of fluctuation diamagnetism in the region of $H_{c2}$, allowing the exact determination of the upper critical field as a function of angle, respectively, in our case, $\theta_{c2}$ as a function of applied field (see inset on the left side of Fig. 8). We then analyzed $H_{c2}(\theta)$ with the Ginzburg-Landau effective mass model to obtain $H_{c2}^{//c}$, $H_{c2}^{//ab}$, and the anisotropy $\gamma$. Although the use of the effective mass model is questionable, as it is incompatible with the apparent temperature dependence of $\gamma$, at constant temperature $H_{c2}(\theta)$ data are well described by it (see left inset of Fig. 8). We note that the scaling procedure used to determine $\theta_{c2}$ contains the target parameter $\gamma$ – therefore, this procedure and the following $H_{c2}(\theta)$ analysis had to be carried out self-consistently.

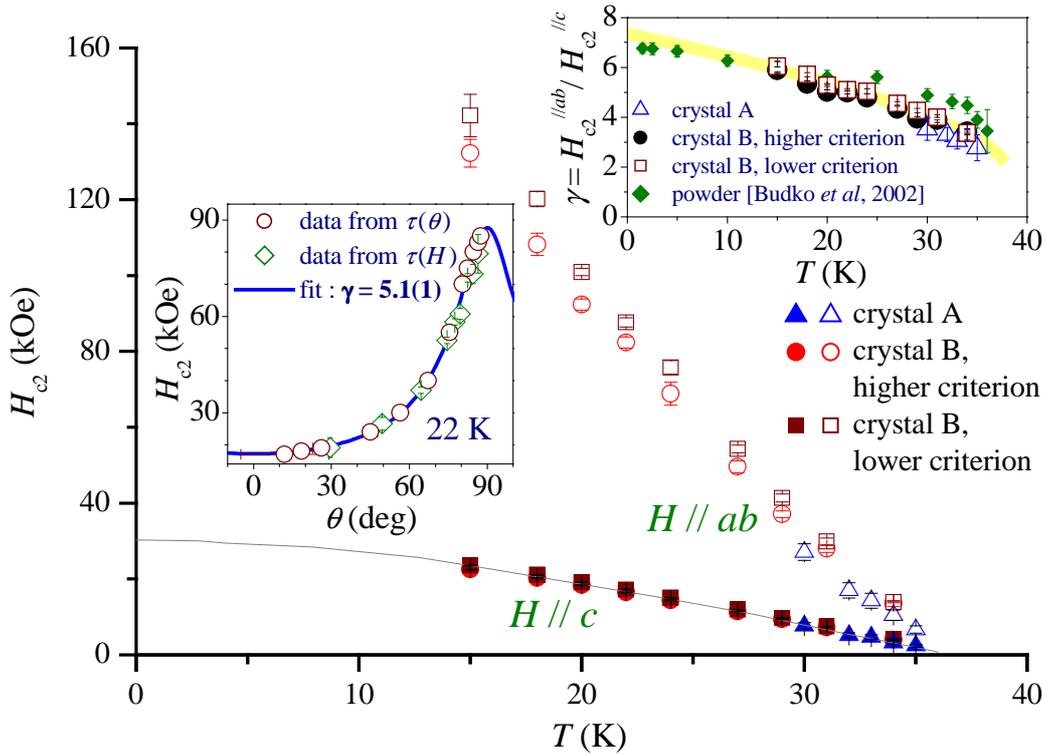

Fig. 8: Upper critical fields parallel (open symbols) and perpendicular (close symbols) to the *ab* plane of the crystal. The different symbols are from measurements of two different crystals and using three different criteria (see Ref. [5]), the full line for H//c corresponds to the usual Helfand-Werthamer [19] dependence. Inset on the left side: Upper critical field $H_{c2}$ as a function of angle $\theta$, at 22 K, from measurements of $\tau$ vs $\theta$ in fixed field $H$ (circles) and $\tau$ vs $H$ at fixed angle $\theta$ (diamonds). Also shown is a fit to the theoretical dependence according to the effective mass model. Upper right inset: $T$ dependence of $\gamma = H_{c2}^{//ab}/H_{c2}^{//c}$. For comparison, later results [20] from an analysis of magnetic measurements of $MgB_2$ powder are also shown.

The upper critical fields resulting from the analysis are shown in the main panel of Fig. 8. It can be seen that the results obtained from two different crystals agree well and that the influence of the exact criteria used to determine $\theta_{c2}$ is not large. The upper right inset of the figure shows the temperature dependence of the $H_{c2}$ anisotropy $\gamma$ corresponding to the data shown in the main panel. Most measurements on single crystals by other techniques, e.g., thermal conductivity [6], specific heat in combination with electrical transport [8,9], or magnetization [7], agree qualitatively well with the results shown in Fig. 8. Quantitative deviations of $H_{c2}^{//ab}(T)$ and $\gamma(T)$ may be due to different defect concentrations in the crystals, since the coherence length can be renormalized by short mean free paths. Additional studies with controlled tuning of defect concentrations are needed to clarify this point.

The strong temperature dependence of the $H_{c2}$ anisotropy is very unusual and it is in disagreement with the predictions of the widely used anisotropic Ginzburg-Landau theory, at least in it's standard form. It can be qualitatively explained by the two different gaps (or order parameters) existing in $MgB_2$, connected with Fermi sheets of different dimensionality. A very recent calculation starting with a phenomenological two band model, but using the most accurate band structure and Eliashberg results [2,3] to determine the effective parameters, yielded values of the upper critical fields and their anisotropy very close to the experimental results shown in Fig. 8 [21].

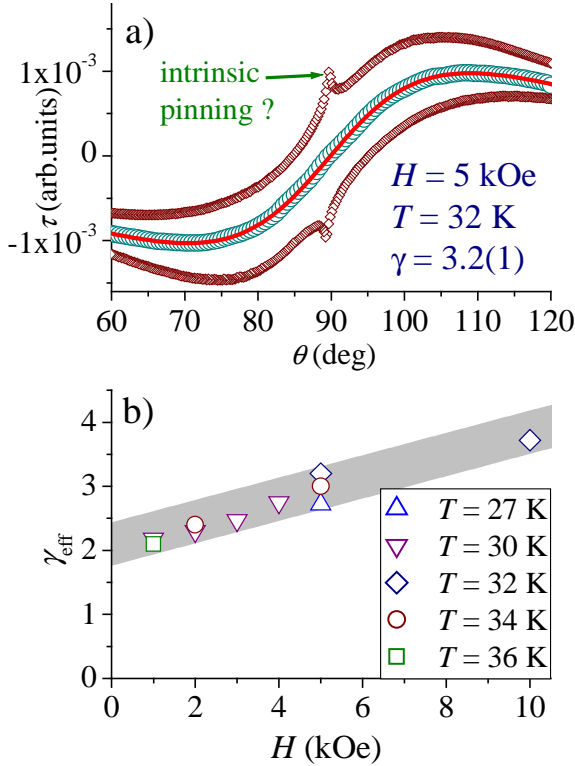

Fig. 9. a) Raw torque $\tau$ vs angle $\theta$ data (diamonds), "shaked" data (circles), and a fit of a theoretical expression based on the anisotropic London model to the "shaked" data (see text). b) Effective anisotropy $\gamma_{eff}$ [as resulting from the analysis of "shaked" $\tau(\theta)$ data with Eq. (1)] vs field $H$, at different temperatures.

We performed also an alternative analysis, using a widely used method to obtain the anisotropy of superconducting state parameters consists of measuring the reversible torque as a function of angle in the mixed state, and analyze the data with an equation, derived first by Kogan *et al*. [22]:

$$\tau = -\frac{\Phi_\circ HV}{64\pi^2\lambda_{ab}^2}\left(1-\frac{1}{\gamma^2}\right)\frac{\sin 2\theta}{\epsilon(\theta)}\ln\left(\frac{\eta H_{c2}^{\|c}}{\epsilon(\theta)H}\right), \quad (1)$$

where $\epsilon(\theta) = (\cos^2\theta + \sin^2\theta/\gamma^2)^{1/2}$, and $\eta$ is a constant of the order of unity. The various symbols in the prefactor are angle independent, for a description see Ref. [23]. Eq. (1) was derived on the basis of the anisotropic London model, valid in the limits of fields $H_{c1} \ll H \ll H_{c2}$. To ensure keeping the latter condition, we restricted the analysis to field directions within 30 degrees from the *ab* planes. A further restriction is that Eq. (1) describes the reversible torque only. When the reversible torque is estimated by simple averaging of the torque values obtained from angle increasing and decreasing scans, usually an analysis with Eq. (1) yields an overestimation of the anisotropy [24], which is the case also in $MgB_2$ [23]. To obtain the true reversible torque, we employed a vortex-shaking process, which speeds up the relaxation of the vortex lattice [24]. As can be seen in Fig. 9a), in the raw data substantial irreversibility is present, but the shaked data are well reversible and can be described by Eq. (1). Figure 9b) shows the result of our analysis with Eq. (1): the so obtained anisotropy shows a pronounced dependence on the applied field, rising roughly linearly from about 2 in zero field to nearly 4 in fields of 10 kOe. In contrast to the upper critical field anisotropy, the anisotropy obtained from the analysis with Eq. (1) seems to be almost temperature independent in the region covered.

We note, however, that the anisotropy obtained from the analysis with Eq. (1) is not necessarily the same as the upper critical field anisotropy, when the standard anisotropic Ginzburg-Landau theory is not applicable (as indicated by the *T* dependence of the $H_{c2}$ anisotropy). This is because in Eq. (1), the anisotropy appears twice – outside of the logarithm, where it is the penetration depth anisotropy $\gamma_\lambda$, and in the logarithm, where it is the upper critical field or coherence length anisotropy $\gamma_\xi$.

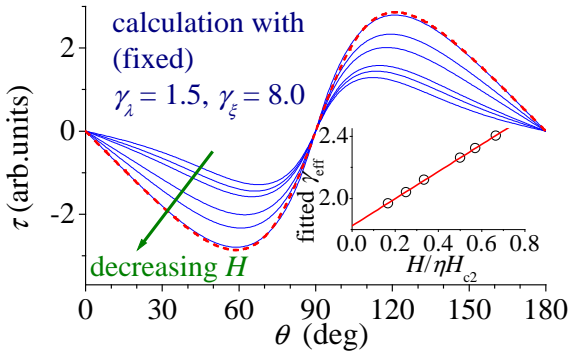

Fig. 10: Torque curves calculated with Eq. (1), but using different (constant) anisotropies for the penetration depth and the coherence length, for various fields (full lines). Also shown is a fit of Eq. (1) using a common effective anisotropy $\gamma_{eff} = \gamma_\lambda = \gamma_\xi$ to one of the calculated curves (broken line). Inset: Field dependence of the effective anisotropy $\gamma_{eff}$ obtained from fits to the calculated curves shown in the main panel.

When these $\gamma_\lambda$ and $\gamma_\xi$ are not the same, the anisotropy obtained from an analysis with Eq. (1) is an effective anisotropy $\gamma_{eff}$. The difference is illustrated in Fig. 10, where torque curves are calculated for different fields, using different, but fixed $\gamma_\lambda = 1.5$ and $\gamma_\xi = 8.0$. Interestingly, for the fitted effective anisotropy becomes field dependent, in a manner similar to the measured data (see inset). Note that a direct fitting of the present data using separate $\gamma_\lambda$ and $\gamma_\xi$ was not possible, since the full formula then is rather ill conditioned numerically. Independent penetration depth measurements [25] indeed found no pronounced anisotropy $\gamma_\lambda$.

Another likely cause of the $H$ dependence of $\gamma_{eff}$ observed in our data is the two-band nature of superconductivity in $MgB_2$. In low fields, the (small gap) 3D $\pi$ sheets of the Fermi surface (FS) exert a heavy influence on the mixed state properties, and dominate, for example, the vortex structure as seen in scanning tunneling spectroscopy [26]. Superconductivity on the $\pi$ sheets of the FS gets, however, strongly suppressed already by moderate fields of the order of 4 kOe [27] and in higher fields superconductivity is largely confined to the 2D $\sigma$ sheets of the FS. Such a crossover from 3D to 2D sheets of the FS should result in an increase of the effective (bulk) anisotropy. Similar conclusions were also drawn from the analysis of specific heat data [27].

In the raw data shown in Fig. 9a), a pronounced peak in the hysteresis is visible for *H//ab*. A similar peak for *H//ab* was observed by other authors as well [28]. It is tempting to ascribe this peak for *H//ab* to "intrinsic pinning" by the non-superconducting layers, analogous to findings in cuprate superconductors. Disturbing is, however, that an estimation [23] based on the superconducting parameters clearly indicates that $MgB_2$ is in the (anisotropic) three-dimensional limit, in which case "intrinsic pinning" is not expected. Measurements on another single crystal showed that the hysteresis peak for *H//ab* is not present in all $MgB_2$ crystals (see Fig. 11) – therefore, it cannot be an intrinsic feature of $MgB_2$. Where present, the effect may be due to a small number of stacking faults, for example. Further details are given in Ref. [23].

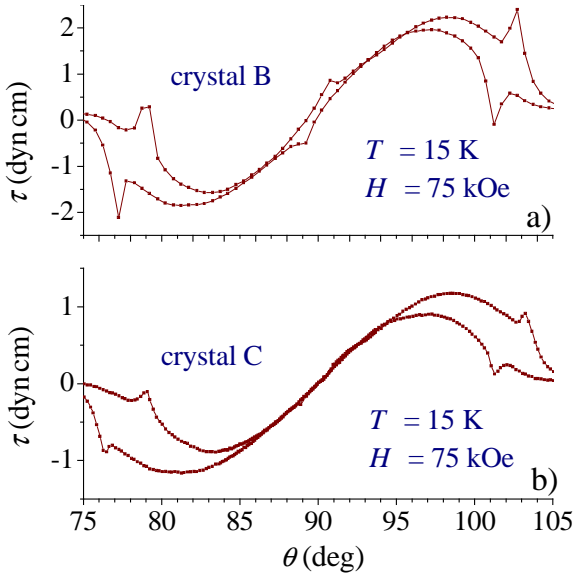

Fig. 11: Torque $\tau$ vs angle $\theta$ at 15 K and 75 kOe for two different crystals. a) Crystal B shows a hysteresis peak for H//ab. b) Crystal C does not show such a peak. Both crystals show a hysteresis peak for angles of about 12 deg between *H* and the *ab* plane. Note that a symmetric (with respect to 90 deg) background was not subtracted.

In contrast, another hysteresis peak, also visible in Fig. 11, was present in all crystals investigated, and thus most likely is of intrinsic origin. This hysteresis peak was found at lower temperatures (below 23 K) in fields of about 0.85 $H_{c2}$. This peak effect (PE) is best investigated with τ(H) measurements at constant angle. An example is shown in the inset of Fig. 12.

We investigated the peak effect region with a minor hysteresis loop (MHL) technique and found pronounced history effects roughly between peak onset and maximum (shaded in the inset of Fig. 12). Details of the MHL study are given in Ref. [29]. We conclude that the peak effect marks a disorder-driven first order phase transition between a Bragg glass and a highly disordered phase with the region between peak onset and maximum being a meta-stability region, where the two phases can co-exist [30]. The vortex matter phase diagram of $MgB_2$, extracted from the torque measurements [29], is shown in Fig. 12. $MgB_2$ has many properties (on the phenomenological level) intermediate between the high $T_c$ cuprate and conventional low $T_c$ superconductors, and detail studies of it's vortex matter phase diagram may help creating a unified description of the phase diagrams of vortex matter for both high and low $T_c$ superconductors.

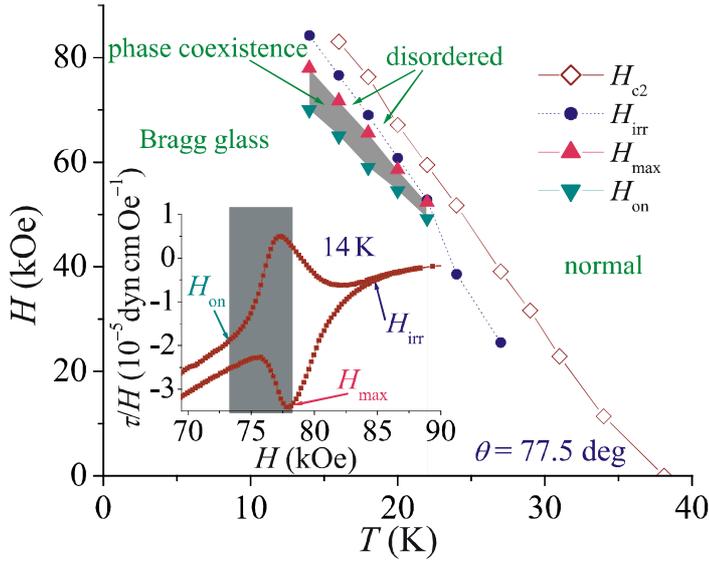

Fig. 12: Field vs temperature phase diagram of $MgB_2$ (crystal B) for an angle θ = 77.5 deg. Shown are the normal phase, the Bragg glass phase, and a highly disordered phase stable between $H_{c2}$ and the peak effect maximum field $H_{max}$. Between $H_{max}$ and the peak onset field $H_{on}$, the disordered phase can coexist with the Bragg glass as a meta-stable phase. Inset: Torque τ/H vs field at 14 K and 77.5 deg in the peak effect region. In the shaded region roughly between peak onset and maximum, the critical current density was found to be history dependent.

## 6. Scanning tunneling spectroscopy (STS) investigations.

Scanning tunneling spectroscopy (STS) measurements were performed on the surface of an as-grown crystal with the tunneling current parallel to the *c* axis. Tunneling in this direction is primarily sensitive to the π-band. This is evident from Fig. 13a), where only the small gap $\Delta \approx 3$ meV connected with π band is observed. The shoulders at ±6 meV are a reminescent of the σ-band. Measurements of STS spectra with the tunneling current parallel to the *ab* plane show a strong signal of a large gap of about 6.9 meV, with a smaller peak of a small gap (about 3.3 meV). Two gaps can thus be observed, with different weights in two directions.

Applying a magnetic field introduces vortices into the sample. These can be imaged, by performing a spatial mapping of the zero bias conductance (ZBC). This way superconducting regions will have a low ZBC, and normal regions a high ZBC. In Fig. 13b). we show the first vortex image obtained on a single crystal of $MgB_2$. The image shows a well ordered hexagonal vortex lattice. Further details are given in Ref. [26].

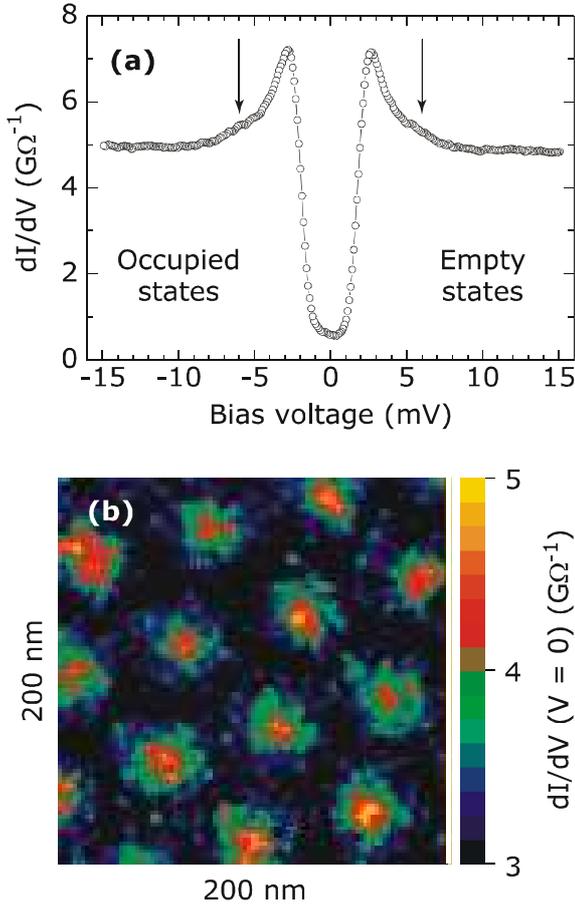

Fig. 13. Results of STS measurements at 1.9 K, with a tunneling resistance of 0.2 GΩ. a) Spectrum obtained in zero field, showing a clear superconducting gap with coherence peaks at ±3 meV. In addition, weak shoulders are visible at ±6 meV (indicated by arrows). b) STS image of the hexagonal vortex lattice induced by an applied field of 5 kOe.

## 7. Decoration experiment

In order to visualise the vortex lattice in $MgB_2$ on a large surface area, a decoration experiment using fine ferromagnetic particles has been performed in a field of 200 Oe. Fig.14 a) shows $MgB_2$ single crystal used for the experiment. Growth steps are visible on a surface of the crystal. Fig.14 b) shows these steps with vortex lattice image. Fig.14 c) shows magnification of vortex lattice image. Vortex lattice has hexagonal symetry.

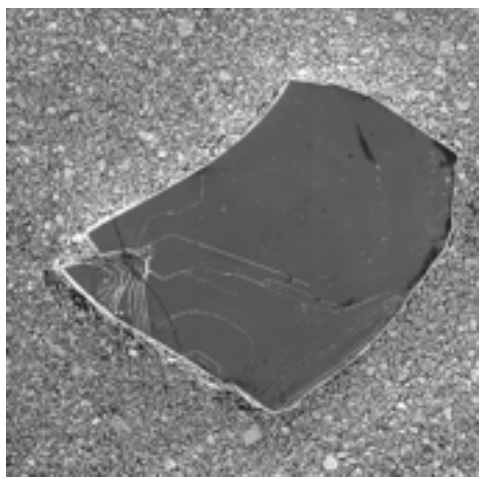
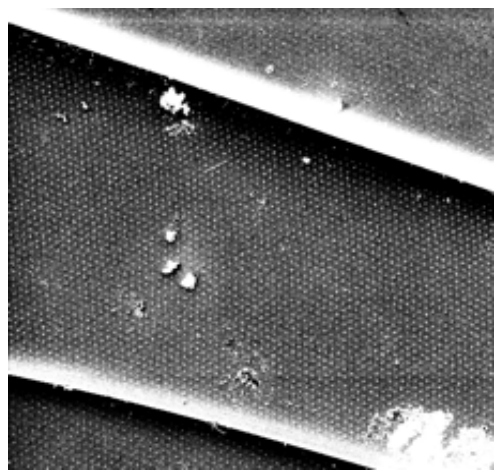

a)                                                   b)

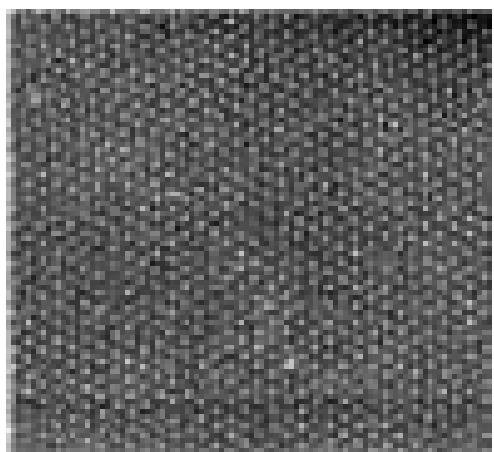

c)

Fig.14. Vortex lattice image obtained with decoration technique. a) Single crystal used for decoration experiment. b) and c) Vortex lattice image. The size of pictures: a) 800 μm, b) 30 μm, c) 15 μm.

## 8. Magneto-optical investigations

We have performed magneto-optical reflectivity measurements in the basal-plane of the hexagonal $MgB_2$ on a mosaic of three $MgB_2$ single crystals with $T_c = 38$ K and a total surface of 2x2 mm$^2$ (one of the crystals used is shown in Fig. 3b). The data were collected from the ultraviolet down to the far infrared as a function of temperature and magnetic field oriented along the c-axis. As shown in Fig. 15, in the far infrared, there is a clear signature of the superconducting gap with a gap-ratio $2\Delta/k_BT_c \sim 1.2$, close to the value expected for the smaller gap on the $\pi$ Fermi sheets. As already mentioned in the previous section, which gap is observed can depend on geometry of the experiment. The gap is fully suppressed only in an external magnetic field of the order of the bulk critical field $H_{c2}(T)$. We have extracted from optical conductivity data this upper critical field $H_{c2}$ along the $c$ axis. The temperature dependence of $H_{c2}^{//c}$ is compatible with single crystal results obtained by other experimental methods (see Sec. 4). A very interesting issue arising from magneto-optical measurements is that these investigations indicate $MgB_2$ to be in the dirty limit, contrary to findings with other experimental techniques. Such a controversy about the clean versus dirty limit scenario awaits resolution. Further details are given in Ref. [32].

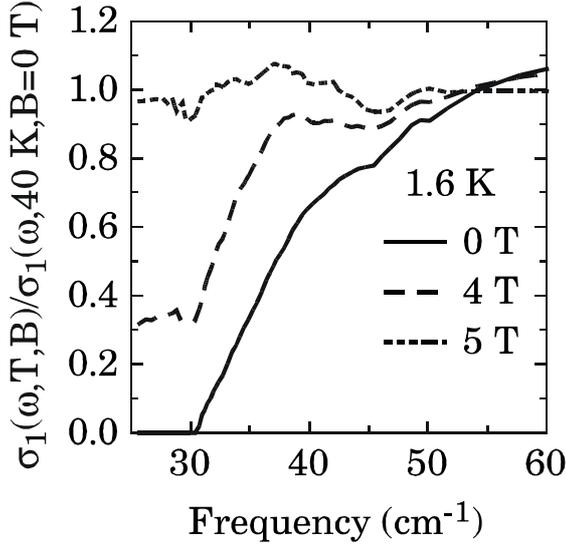

Fig. 15. Magnetic field dependence (*H*//*c*) of the conductivity ratio $\sigma_{1s}/\sigma_{1n}$ at 1.6 K in the far infrared spectral range. The normal state $\sigma_{1n}$ corresponds to the measurement at 40 K in zero field.

**9. Conclusions**

Single crystals of $MgB_2$ of millimeter size have been grown, using a high pressure anvil technique. $MgB_2$ crystals grow by a peritectic decomposition reaction of the intermediate nitride $MgNB_9$. X-ray investigations show a full occupation of atomic positions and low values of agreement factor R. This, together with the sharp transitions to the superconducting state observed in magnetic measurements, proves the very high quality of crystals.

Investigations with torque magnetometry revealed a temperature and field dependence of the anisotropy of $MgB_2$. This dependence can be explained in a two band, two gap model. Our torque results imply a breakdown of standard anisotropic Ginzburg-Landau theory with a constant effective mass anisotropy. A pronounced peak effect, with accompanying history dependent critical current density, was also observed by torque magnetometry. The peak marks a disorder-driven first order phase transition between a Bragg glass and a highly disordered vortex phase. A *H,T* phase diagram of the vortex matter of $MgB_2$ is proposed.

Scanning tunneling spectra show gap signatures at two energies, depending on the tunneling direction, directly showing the existence of two energy gaps of different dimensionality. Decoration experiments show the vortex lattice on the whole surface of a single crystal.

For the first time, magneto-optic investigations on single crystals allowed observation of the small surperconducting gap in $MgB_2$ and it's depression by an applied magnetic field.

**Acknowledgments**

The authors acknowledge support from the Swiss National Science Foundation, by the European Community (program ICA1-CT-2000-70018, Centre of Excellence CELDIS) and by the Polish State Committee for Scientific Research (KBN, Contract No.5 P03B 12421).